\documentclass[a4paper,10pt]{article}
\usepackage[utf8x]{inputenc}

\usepackage{color}

\usepackage[normalem]{ulem}

\def\Comment#1{}
\newcommand{\bean}{\begin{eqnarray*}}
\newcommand{\eean}{\end{eqnarray*}}

\newcommand{\gapproxeq}{\lower
.7ex\hbox{$\;\stackrel{\textstyle >}{\sim}\;$}}
\newcommand{\lapproxeq}{\lower
.7ex\hbox{$\;\stackrel{\textstyle <}{\sim}\;$}}

\newcommand\lsim{\mathrel{\rlap{\lower4pt\hbox{\hskip1pt$\sim$}}
    \raise1pt\hbox{$<$}}}
\newcommand\gsim{\mathrel{\rlap{\lower4pt\hbox{\hskip1pt$\sim$}}
    \raise1pt\hbox{$>$}}}
\newcommand{\ba}{\begin{array}}
\newcommand{\ea}{\end{array}}

\newcommand{\be}{\begin{small}\begin{equation}}
\newcommand{\ee}{\end{equation}\end{small}}
\newcommand{\bear}{\begin{small}\begin{eqnarray}}
\newcommand{\eear}{\end{eqnarray}\end{small}}

\newcommand{\ket}{\,\rangle}
\newcommand{\bra}{\langle \,}

\newcommand{\cO}{{\cal O}}

\newcommand{\mL}{\mathcal{L}}

\newcommand{\mF}{\mathcal{F}}

\newcommand{\mO}{\mathcal{O}}

\def\bat{\begin{array}{cc}}

\newcommand{\Frac}[2]{\frac{\displaystyle #1}{\displaystyle #2}}

\newcommand{\bea}{\begin{small}\begin{eqnarray}}
\newcommand{\eea}{\end{eqnarray}\end{small}}

\begin{document}

\rightline{\large FTUAM-13-38}
\rightline{\large IFT-UAM/CSIC-13-136}

\vspace*{2.0cm}

\begin{center}
{\LARGE \textbf{
Consistent high-energy constraints \\
in the anomalous QCD sector
\\[2cm]}}

{\large \textbf{Pablo Roig$^{1}$ and Juan Jos\'e Sanz Cillero$^{2}$}\\[1.2 cm]}
 $\ ^{1}$\textit{Instituto de F\'{\i}sica, Universidad Nacional Aut\'onoma de M\'exico,
AP 20-364, M\'exico D.F. 01000, M\'exico}\\[0.4cm]
$\ ^{2}$\textit{Departamento de F\'{\i}sica Te\'orica and Instituto de F\'{\i}sica Te\'orica, IFT-UAM/CSIC, Universidad
Aut\'onoma de Madrid, Cantoblanco, E-28049 Madrid, Spain}
\end{center}

\vspace*{2.0cm}
\begin{abstract}
The anomalous $\bra VVP\ket $ Green function
and related form-factors ($\pi^0\to \gamma^*\gamma^*$
and $\tau^-\to X^-\nu_\tau$ vector form-factors, with
$X^-=(KK\pi)^-,\, \varphi^-\gamma ,\, (\varphi V)^-$)
are analyzed in this letter in the large--$N_C$ limit.
Within the single (vector and pseudoscalar) resonance
approximation and the context of Resonance Chiral Theory,
we show that all these observables over-determine in a consistent way
a unique set of compatible high-energy constraints for the resonance couplings.
This result is in agreement with analogous relations found in the even
intrinsic-parity sector of QCD like, e.g., $F_V^2 = 3 F^2$.
The antisymmetric tensor formalism is considered for the spin-one resonance fields.
Finally, we have also worked out and provide here the relation between the two
bases of odd intrinsic-parity Lagrangian operators commonly employed in the literature.

\end{abstract}
\vspace*{2.0cm}

\section{Introduction}

Chiral symmetry plays a crucial role in the structure of
non-perturbative light-quark interactions in Quantum Chromodynamics (QCD).
However, it becomes spontaneously broken, generating the corresponding
chiral (pseudo) Goldstones $\varphi^a$. Its low-energy interaction
can be then described through an effective field theory (EFT) which
implements chiral symmetry and the discrete symmetries of QCD, denoted as
Chiral Perturbation Theory
($\chi$PT)~\cite{Weinberg:1978kz, Gasser:1983yg, Gasser:1984gg, Bijnens:1999sh, Bijnens:2001bb}.
The Wess-Zumino-Witten (WZW) term
reproduces the chiral anomaly of QCD~\cite{Wess:1971yu, W}
and will provide the leading contribution to the anomalous QCD amplitudes
at low energies.

  Resonance Chiral Theory (R$\chi$T)~\cite{Ecker:1988te, Ecker:1989yg}
describes the interactions between the light-quark resonances
and the chiral pseudo-Goldstones based on chiral invariance and using
the formal expansion in $1/N_C$~\cite{'tHooft:1973jz, 'tHooft:1974hx, Witten:1979kh}
as a guiding principle to sort out perturbative computations,
with $N_C$ the number of colours in QCD.
In most applications, one is forced to truncate the resonance spectrum and
just the lightest resonance multiplets are included~\footnote{
The truncation of the infinite tower of large--$N_C$ resonances introduces
in general a theoretical uncertainty in our determinations and
may lead to some issues when a broader and broader set of observables is
analyzed~\cite{truncation-Prades,truncation}.}.

The R$\chi$T Lagrangian is built in such a way that it fulfills
the discrete symmetries of QCD and chiral symmetry.
It contains an even~\cite{Ecker:1988te,Ecker:1989yg,Cirigliano:2006hb}
and an odd intrinsic-parity sector~\cite{RuizFemenia:2003hm,Kampf:2011ty},
where the latter produces contributions to anomalous amplitudes in QCD.
{Nonetheless, it should be pointed out that, apart from the WZW term, the R$\chi$T interactions
are chiral invariant.}

Although QCD fixes the hadronic couplings, these cannot be determined
based on chiral symmetry considerations alone. {\it A priori}, they are free parameters
of our R$\chi$T action and chiral symmetry just provides relations between
particular processes. However, demanding a short-distance behaviour
in accordance with QCD and its Operator Product Expansion~(OPE)~\cite{SVZ}
will allow us to extract important constraints between couplings like, {\it e.g.},
the Weinberg Sum Rules (WSRs)~\cite{WeinbergSRS}.

Along the last years, there has been an extensive program of computation
of Green functions and associated form-factors within the R$\chi$T
framework~\cite{Ecker:1988te, Ecker:1989yg, truncation-Prades,Cirigliano:2006hb, RuizFemenia:2003hm,Kampf:2011ty, Knecht:2001xc, EarlierWorks, Cirigliano:2004ue,
Cirigliano:2005xn}, where a key ingredient has been the implementation
of the high-energy QCD behaviour prescribed by the OPE~\cite{SVZ,Mateu:2008}
and the quark-counting rules for hadronic form-factors~\cite{Brodsky:1973kr}.

The difficulties to match the anomalous $\bra VVP \ket$
Green function (between two vector currents and one
pseudoscalar density) in the four-vector
field representation for spin-one resonances~\cite{truncation-Prades,Knecht:2001xc,EarlierWorks}
triggered the research reported in Ref.~\cite{RuizFemenia:2003hm}.
Therein, the authors included the chiral pseudo-Goldstones
and the lightest vector multiplet and showed that these issues could be solved
in the antisymmetric tensor field formalism for the vector
resonances~\cite{Ecker:1988te, Ecker:1989yg}.
This will be the formalism used all along this letter.

As result of that $\bra VVP\ket$ study~\cite{RuizFemenia:2003hm},
there were several subsequent analyses~\cite{Cirigliano:2004ue, Cirigliano:2005xn}
where other Green functions that are order parameters of chiral symmetry breaking  were evaluated
in an analogous way.
Finally, a global understanding of the
short-distance constraints on these Green functions and related form-factors
was presented in Ref.~\cite{Cirigliano:2006hb},
together with an exhaustive evaluation of the
resonance contributions to the $\chi$PT low-energy constants (LECs)
of the even intrinsic-parity sector up to $\mathcal{O}(p^6)$.

However, the analysis of $\tau^-\to (KK\pi)^- \nu_\tau$ decays \cite{Dumm:2009kj}
revealed a set of high-energy constraints on the R$\chi$T couplings
which was at odds with the findings of Ref.~\cite{RuizFemenia:2003hm}.
Furthermore, an inconsistent asymptotic behaviour
for the $\pi^0\to\gamma\gamma^*$ pion transition form-factor (TFF)
was found~\cite{Dumm:2009kj} if the high-energy
restrictions~\cite{RuizFemenia:2003hm} were considered.
After that, Ref.~\cite{Kampf:2011ty} revisited the $\left\langle VVP\right\rangle$
Green function. It extended the odd intrinsic-parity
resonance Lagrangian, allowing for operators with multiple resonance fields,
verified previous analyses of the saturation of  odd intrinsic-parity
$\mathcal{O}(p^6)$ LECs under integration of the resonances
and performed a complete study of the saturation to the full set
of odd $\cO(p^6)$ LECs~\cite{Bijnens:1999sh,Bijnens:2001bb}.
On the contrary to Ref.~\cite{RuizFemenia:2003hm},
the short-distance behaviours of the pion TFF and the $\bra VVP \ket$ Green function were found to be
compatible with QCD, provided the lightest pseudoscalar resonance multiplet
was also taken into account (it was not included in the previous work~\cite{RuizFemenia:2003hm}).
Since the basis of Lagrangian operators employed in both
references~\cite{RuizFemenia:2003hm,Kampf:2011ty} was different,
it still remained unclear whether all inconsistencies among
the referred high-energy constraints
were fully solved by adding a pseudoscalar resonance multiplet to the action or not.

The aim of this letter is to answer this question. We will show that
the odd intrinsic-parity R$\chi$T Lagrangian including
the pseudo-Goldstones and the lightest multiplet of pseudoscalar
and vector resonances~\cite{Kampf:2011ty}
produces a unique consistent set of short-distance relations
for the $\left\langle VVP\right\rangle$
Green function and the associated form-factors studied
so far in the literature,
in an analogous way to the agreement
found in the even intrinsic-parity
sector~\cite{Cirigliano:2006hb}.

\vspace*{0.25cm}
\section{The odd intrinsic-parity resonance Lagrangian}

 The operators in the R$\chi$T Lagrangian can be classified according
to the number of resonance fields:
\bear
\mL_{R\chi T} &=& \mL_G\, +\, \sum_R \mL_R\, +\, \sum_{R,\, R'} \mL_{RR'}\, +\, ...
\eear
with $\mL_G$ given by the operators with only pseudo-Goldstone fields
and external vector and axial-vector sources.
It contains the even-parity $\cO(p^2)$ $\chi$PT
Lagrangian~\cite{Gasser:1983yg,Gasser:1984gg,Ecker:1988te}
and the WZW term~\cite{Wess:1971yu,W,RuizFemenia:2003hm,Kampf:2011ty}.

At leading order in the $1/N_C$ expansion,
the most general odd intrinsic-parity resonance chiral Lagrangian
for processes involving one pseudo-Goldstone and two vector objects
(two vector resonances, or one external source and one vector resonance)
was derived in Ref.~\cite{RuizFemenia:2003hm}:
\begin{equation}\label{eq: Lagrangian Pedro}
\mathcal{L}_{R\chi T}^{\mathrm{odd}}\,
\supset \,\sum_{i=1}^{7} \,\frac{c_i}{M_{V}}\,{\cal O}^i_{\mbox{\tiny{VJP}}} \; + \;
\sum_{i=1}^{4} \,d_i\,{\cal O}^i_{\mbox{\tiny{VVP}}}\; .
\end{equation}
The monomials ${\cal O}^i_{\mbox{\tiny{VJP}}}$ and
${\cal O}^i_{\mbox{\tiny{VVP}}}$ are
provided in Table~\ref{tab:1-Portoles}.
The antisymmetric tensor formulation is employed here to describe
the spin--1 fields~\cite{Ecker:1988te, Ecker:1989yg}.
Analogous analyses with the spin--1 fields given in the four-vector (Proca)
formalism can be found in Ref.~\cite{truncation-Prades,EarlierWorks}, although the present work
only studies the antisymmetric tensor representation.

We follow the notation and conventions of Ref.~\cite{Cirigliano:2006hb},
where the chiral tensors entering $\mathcal{L}_{R\chi T}^{\mathrm{odd}}$
are defined.
$M_V$ is the vector resonance multiplet mass
in the chiral and large--$N_C$ limits.

Within the same framework, and motivated by the analogous work
on the even-intrinsic parity resonance Lagrangian accomplished
in Ref.~\cite{Cirigliano:2006hb}, the authors of
Ref.~\cite{Kampf:2011ty} constructed the most general resonance chiral Lagrangian
in the odd intrinsic-parity sector that can generate chiral low-energy constants
up to $\mathcal{O}(p^6)$ \cite{Bijnens:1999sh, Bijnens:2001bb}.
They considered the contribution from the lightest resonance multiplets,
in particular vector ($V$) and pseudoscalar ($P$) resonances.
The latter was absent in the previous treatment in Ref.~\cite{RuizFemenia:2003hm}.
The part of the odd intrinsic-parity Lagrangian relevant
for the $\bra VVP\ket$ Green function and the related form-factors studied
in this article is given by~\cite{Kampf:2011ty}
\begin{equation}\label{eq: Lagrangian Kampf}
\tilde{\mathcal{L}}_{R\chi T}^{\mathrm{odd}}\,\supset
\,\sum_i\sum_X \kappa_i^X \varepsilon^{\mu\nu\alpha\beta}\widehat{\mathcal{O}}_{i\;\mu\nu\alpha\beta}^X\; ,\quad X\;=\;V,\,VV,\,PV\,,
\end{equation}
where the corresponding operators can be read in Tables \ref{tab:1} and \ref{tab:2}.

\begin{table}[!t]
\begin{center}
\begin{tabular}{|c|c||c|c|}
\hline
$\phantom{\bigg|}$
$i$ & $ {\cal O}_{\mbox{\tiny VJP}}^i$
& $i$ & ${\cal O}_{\mbox{\tiny VVP}}^i$
\\ \hline
1 & $\varepsilon_{\mu\nu\rho\sigma}\,
\langle \, \{V^{\mu\nu},f_{+}^{\rho\alpha}\} \nabla_{\alpha}u^{\sigma}\,\rangle
$
& 1 & $ \varepsilon_{\mu\nu\rho\sigma}\,
\langle \, \{V^{\mu\nu},V^{\rho\alpha}\} \nabla_{\alpha}u^{\sigma}\,\rangle $
\\ \hline
2 & $ \varepsilon_{\mu\nu\rho\sigma}\,
\langle \, \{V^{\mu\alpha},f_{+}^{\rho\sigma}\} \nabla_{\alpha}u^{\nu}\,\rangle$
& 2 & $i\,\varepsilon_{\mu\nu\rho\sigma}\,
\langle \, \{V^{\mu\nu},V^{\rho\sigma}\}\, \chi_{-}\,\rangle$
\\ \hline
3 & $i\,\varepsilon_{\mu\nu\rho\sigma}\,
\langle \, \{V^{\mu\nu},f_{+}^{\rho\sigma}\}\, \chi_{-}\,\rangle$
& 3 & $\varepsilon_{\mu\nu\rho\sigma}\,
\langle \, \{\nabla_{\alpha}V^{\mu\nu},V^{\rho\alpha}\} u^{\sigma}\,\rangle$
\\ \hline
4 & $i\,\varepsilon_{\mu\nu\rho\sigma}\,
\langle \, V^{\mu\nu}\,[\,f_{-}^{\rho\sigma}, \chi_{+}]\,\rangle$
& 4 & $\varepsilon_{\mu\nu\rho\sigma}\,
\langle \, \{\nabla^{\sigma}V^{\mu\nu},V^{\rho\alpha}\} u_{\alpha}\,\rangle$
\\ \hline
5 & $ \varepsilon_{\mu\nu\rho\sigma}\,
\langle \, \{\nabla_{\alpha}V^{\mu\nu},f_{+}^{\rho\alpha}\} u^{\sigma}\,\rangle$
&   &
\\ \hline
6 & $\varepsilon_{\mu\nu\rho\sigma}\,
\langle \, \{\nabla_{\alpha}V^{\mu\alpha},f_{+}^{\rho\sigma}\} u^{\nu}\,\rangle$
&   &
\\ \hline
7 & $\varepsilon_{\mu\nu\rho\sigma}\,
\langle \, \{\nabla^{\sigma}V^{\mu\nu},f_{+}^{\rho\alpha}\} u_{\alpha}\,\rangle$
&   &
\\ \hline
\end{tabular}%
\end{center}
\caption{\small
Monomials with one vector resonance field ($\cO^i_{VJP}$) and
two vector fields ($\cO^i_{VVP}$) in the basis of Ref.~\cite{RuizFemenia:2003hm}.}
\label{tab:1-Portoles}
\end{table}

\begin{table}[!t]
\begin{center}
\begin{tabular}{|c|c||c|c|}
\hline
$\phantom{\bigg|}$
$i$ & $\widehat{\mathcal{O}}^V_{i\,\mu\nu\alpha\beta}$ & $i$ & $\widehat{%
\mathcal{O}}^V_{i\,\mu\nu\alpha\beta}$ \\ \hline
1 & $\mathrm{i} {\langle} V_{\mu\nu}(h_{\alpha\sigma}u^\sigma u_\beta -
u_\beta u^\sigma h_{\alpha\sigma}){\rangle}$ & 11 & ${\langle}
V_{\mu\nu}\{f_{+\,\alpha\rho},f_{-\,\beta\sigma}\}{\rangle} g^{\rho\sigma}$ \\
\hline
2 & $\mathrm{i}{\langle} V_{\mu\nu}(u^\sigma h_{\alpha\sigma} u_\beta -
u_\beta h_{\alpha\sigma} u^\sigma)$ & 12 & ${\langle} V_{\mu\nu}
\{f_{+\,\alpha\rho},h_{\beta\sigma}\}{\rangle} g^{\rho\sigma}$ \\ \hline
3 & $\mathrm{i}{\langle} V_{\mu\nu}(u^\sigma u_\beta h_{\alpha\sigma} -
h_{\alpha\sigma}u_\beta u^\sigma){\rangle}$ & 13 & $\mathrm{i}{\langle}
V_{\mu\nu}f_{+\,\alpha\beta}{\rangle}{\langle} \chi_-{\rangle}$ \\ \hline
4 & $\mathrm{i}{\langle} [V_{\mu\nu},\nabla_\alpha\chi_+]u_\beta{\rangle} $
& 14 & $\mathrm{i} {\langle} V_{\mu\nu}\{f_{+\,\alpha\beta},\chi_-\}{\rangle}$
\\ \hline
5 & $\mathrm{i}{\langle} V_{\mu\nu}[f_{-\,\alpha\beta},u_\sigma u^\sigma]{%
\rangle}$ & 15 & $\mathrm{i} {\langle} V_{\mu\nu}[f_{-\,\alpha\beta},\chi_+]{%
\rangle}$ \\ \hline
6 & $\mathrm{i}{\langle} V_{\mu\nu}(f_{-\,\alpha\sigma}u_\beta u^\sigma -
u^\sigma u_\beta f_{-\,\alpha\sigma}){\rangle}$ & 16 & ${\langle}
V_{\mu\nu}\{\nabla_\alpha f_{+\,\beta\sigma},u^\sigma\}{\rangle}$ \\ \hline
7 & $\mathrm{i}{\langle} V_{\mu\nu}(u^\sigma f_{-\,\alpha\sigma}u_\beta -
u_\beta f_{-\,\alpha\sigma}u^\sigma){\rangle}$ & 17 & ${\langle}
V_{\mu\nu}\{\nabla^\sigma f_{+\,\alpha\sigma},u_\beta\}{\rangle}$ \\ \hline
8 & $\mathrm{i} {\langle} V_{\mu\nu}(f_{-\,\alpha\sigma} u^\sigma u_\beta -
u_\beta u^\sigma f_{-\,\alpha\sigma}){\rangle} $ & 18 & ${\langle} V_{\mu\nu}
u_\alpha u_\beta {\rangle} {\langle} \chi_-{\rangle}$ \\ \hline
9 & ${\langle} V_{\mu\nu}\{\chi_-,u_\alpha u_\beta\}{\rangle} $ &  &  \\
\hline
10 & ${\langle} V_{\mu\nu} u_\alpha \chi_- u_\beta {\rangle}$ &  &  \\ \hline
\end{tabular}%
\end{center}
\caption{\small
Monomials with one vector resonance field in the basis of Ref.~\cite{Kampf:2011ty}.}
\label{tab:1}
\end{table}

\begin{table}[!t]
\begin{center}
\begin{tabular}{|c|c||c|c|}
\hline
$\phantom{\bigg|}$
$i$ &
$\widehat{\mathcal{O}}^{VV}_{i\,\mu\nu\alpha\beta}$& $i$ &
$\widehat{\mathcal{O}}^{PV}_{i\,\mu\nu\alpha\beta}$ \\ \hline\hline
1 & $\mathrm{i}\langle V_{\mu\nu}V_{\alpha\beta}\rangle \langle \chi_- \rangle $ &
1 & $\mathrm{i} {\langle} \{V_{\mu\nu},P\}u_\alpha u_\beta{\rangle} $\\ \hline
2 & $\mathrm{i}\langle \{V_{\mu\nu},V_{\alpha\beta}\}\chi_-\rangle$ &
2 &  $\mathrm{i}{\langle} V_{\mu\nu} u_\alpha P u_\beta{\rangle}$\\ \hline
3 & $\langle \{\nabla^\sigma V_{\mu\nu},V_{\alpha\sigma}\} u_\beta \rangle$ &
3 &  ${\langle}\{V_{\mu\nu},P\} f_{+\,\alpha\beta}{\rangle}$\\ \hline
4 & $\langle \{\nabla_\beta V_{\mu\nu},V_{\alpha\sigma}\}u^\sigma \rangle$ & & \\ \hline
\end{tabular}%
\end{center}
\caption{\small
Monomials with two vector resonance fields (left hand side)
and a pseudoscalar resonance and a vector resonance field (right hand side)
in the basis of Ref.~\cite{Kampf:2011ty}.}
\label{tab:2}
\end{table}

It is possible to rewrite the resonance Lagrangian from
Ref.~\cite{RuizFemenia:2003hm} (Table~\ref{tab:1-Portoles})
in terms of the basis of monomials in
Ref.~\cite{Kampf:2011ty} (Tables~\ref{tab:1} and~\ref{tab:2})
by means of partial integration and the Bianchi and Schouten
identities~\cite{Bijnens:2001bb}.
This exercise yields the following relations between
the $\tilde{\mL}^{\rm odd}_{R\chi T}$ and the ${\mL}^{\rm odd}_{R\chi T}$
couplings:
\begin{eqnarray}\label{eq: relation different basis}
& & \kappa_1^{VV}= \frac{-d_1}{8 n_f}\,,\qquad \kappa_2^{VV}
= \frac{d_1}{8} + d_2\,,\qquad \kappa_3^{VV}=d_3\,,\qquad
\kappa_4^{VV}=d_4\,,
\nonumber\\
& &
-2 M_V \kappa_5^V \, = \,  M_V \kappa_6^V \, =\, M_V \kappa_7^V = \frac{c_6}{2}\,,
\qquad \qquad
M_V \kappa_{11}^V = \frac{c_1 - c_2 - c_5 + c_6 + c_7}{2}\,,
\nonumber\\
& & M_V \kappa_{12}^V = \frac{c_1 - c_2 - c_5 + c_6 - c_7}{2}\,,
\qquad n_f M_V \kappa_{13}^V = \frac{- c_2 + c_6}{4}\,,
\qquad M_V \kappa_{14}^V = \frac{c_2 + 4 c_3 - c_6}{4}\,,\nonumber\\
& &
 M_V \kappa_{15}^V = c_4\,,
\qquad
M_V \kappa_{16}^V = c_6 + c_7\,,
\qquad
M_V \kappa_{17}^V = -c_5 + c_6\,.
\label{eq.dictionary}
\end{eqnarray}
No high-energy constraint is considered for the derivation
of these relations.

The present study of the anomalous sector at high energies
also requires the following pieces of the even-intrinsic
parity Lagrangian~\cite{Ecker:1988te}:
\begin{equation}\label{eq: original L}
 \mathcal{L}_{R\chi T}^{\mathrm{even}}\supset \frac{F_V}{2\sqrt{2}} \left\langle V_{\mu\nu} f_+^{\mu\nu}\right\rangle\,+\,i\,d_m \left\langle P \chi_-\right\rangle\,.
\end{equation}

\vspace*{0.25cm}
\section{High-energy constraints}

\vspace*{0.25cm}
\subsection{$\bra VVP \ket$ Green function}
\label{sec.VVP}

 We consider first the Green function $\bra VVP\ket$ between
two vector currents $J_V^{\mu,a}(x)$ and $J_V^{\nu,b}(y)$
and one pseudoscalar density $J_P^c(z)$. In momentum space, the OPE prescribes
a very precise short-distance behaviour for
$\Pi^{\mu\nu}_{VVP}(\lambda p,\lambda q,\lambda r)$
when $\lambda\to \infty$~\cite{Mateu:2008}.
Matching the $\bra VVP\ket$ Green function prediction from the resonance
chiral Lagrangian $\mathcal{L}_{R\chi T}^{\mathrm{odd}}$
in eq.~(\ref{eq: Lagrangian Pedro})
and the previously referred OPE asymptotic behaviour
yields \cite{RuizFemenia:2003hm}
\begin{eqnarray}
4 \, c_3+c_1&=&0 \; \; \; ,
\label{eq: constraints PedroA}
\\[3mm]
c_1-c_2+c_5 &=& 0 \; \; \; ,
\label{eq: constraints PedroB}
\\[3mm]
c_5-c_6&=& \frac{N_C}{64\pi^2}\frac{M_V}{\sqrt{2}F_V} \; \; \; ,
\label{eq: constraints PedroC}
\\[3mm]
d_1 + 8 \, d_2&=& -\frac{N_C}{64\pi^2}\frac{M_V^2}{F_V^2} \, +
\, \frac{F^2}{4F_V^2}  \; \; \; ,
\label{eq: constraints PedroD}
\\[3mm]
d_3&=&-\frac{N_C}{64\pi^2}\frac{M_V^2}{F_V^2} \, + \, \frac{F^2}{8F_V^2}\,.
\label{eq: constraints Pedro}
\label{eq: constraints PedroE}
\end{eqnarray}

If one incorporates pseudoscalar resonances and the $\hat{\mO}^{PV}_i$
operators from Lagrangian $\tilde{\mathcal{L}}_{R\chi T}^{\mathrm{odd}}$
in eq.~(\ref{eq: Lagrangian Kampf}), one now
obtains~\cite{Kampf:2011ty}~\footnote{We omit the
prediction for the coupling of another operator which involves
just one pseudoscalar resonance field, as it does not affect our discussion.}
\begin{eqnarray}
M_V( 2 \kappa_{12}^V + 4 \kappa_{14}^V +\kappa_{16}^V -\kappa_{17}^V) \quad =&
4 \,c_3 \,+\, c_1 & = \quad  0\,,
\label{eq.constraint-Kampf-A}
\\[3mm]
M_V (2\kappa_{12}^V + \kappa_{16}^V -2\kappa_{17}^V) \quad =&
c_1 \, - \, c_2 \, + \, c_5 \, & = \quad  0 \, ,
\label{eq.constraint-Kampf-B}
\\[3mm]
-\, M_V \kappa_{17}^V \quad = &  c_5 \, - \, c_6 \, & =
\quad  \frac{N_C \,M_V}{64 \,\sqrt{2}\, \pi^2\,F_V} \, ,
\label{eq.constraint-Kampf-C}
\\[3mm]
8\kappa_2^{VV} \quad =&
d_1 + 8 \, d_2&=\quad  -\frac{N_C}{64\pi^2}\frac{M_V^2}{F_V^2} \, +
\, \frac{F^2}{4F_V^2}
\,+\, \Frac{4 \sqrt{2} d_m \kappa_3^{PV}}{F_V} \; \; \; ,
\label{eq.constraint-Kampf-D}
\\[3mm]
\kappa_3^{VV}\quad = & d_3&=\quad
-\frac{N_C}{64\pi^2}\frac{M_V^2}{F_V^2} \, + \, \frac{F^2}{8F_V^2}
\,+\, \Frac{4 \sqrt{2} d_m \kappa_3^{PV}}{F_V}\,.
\label{eq.constraint-Kampf-E}
\end{eqnarray}
The five constraints derived in Ref.~\cite{Kampf:2011ty} for
$\kappa_{14}^V$, $(2\kappa_{12}^V+\kappa_{16}^V)$, $\kappa_{17}^V$, $\kappa_2^{VV}$
and $(8\kappa_2^{VV}-\kappa_3^{VV})$ have been recast
in the equivalent form in
eqs.~(\ref{eq.constraint-Kampf-A})--(\ref{eq.constraint-Kampf-E}).
The $\kappa_l^{V}$ and $\kappa_m^{VV}$ couplings have been rewritten in terms
of the $c_i$ and $d_j$ couplings
of the $\mL^{\rm odd}_{R\chi T}$ Lagrangian by means of the relations in
eq.~(\ref{eq.dictionary}).
This reproduces the first three constraints derived from
$\mL^{\rm odd}_{R\chi T}$ in
eqs.~(\ref{eq: constraints PedroA})--(\ref{eq: constraints PedroC}).
Notice, however, that the inclusion of the lightest pseudoscalar resonance multiplet modifies
the constraints~(\ref{eq: constraints PedroD})
and~(\ref{eq: constraints Pedro}).

\vspace*{0.25cm}
\subsection{$\tau^- \to (KK\pi)^- \nu_\tau$ form-factors}

A series of high-energy constraints were extracted from the analysis of the
$\tau^-\to (KK\pi)^-\nu_\tau$ decays~\cite{Dumm:2009kj}
after demanding that the corresponding
contribution to the spectral function of the vector-vector correlator
vanished asymptotically~\footnote{
The two-point Green functions of vector and axial-vector currents were studied
within perturbative QCD in Ref.~\cite{Floratos:1978jb},
where it was shown that both spectral functions go to a constant value at infinite momentum transfer.}:
\begin{eqnarray}
\label{eq:1c}
M_V(2 \kappa_{12}^V +\kappa_{16}^V - 2\kappa_{17}^V) \quad =&
c_1 \, - \, c_2 \, + \, c_5 \, & = \quad  0 \, ,
\\
\label{eq:2c}
-\, M_V\kappa_{17}^V \quad =& c_5-c_6 &=\quad \Frac{N_C}{ 192  \pi^2}\Frac{F_V M_V}{\sqrt{2} F^2}\, ,
\\
\label{eq:3c}
\kappa_3^{VV}\quad = & d_3 & =\quad
- \frac{N_C}{192 \, \pi^2} \, \frac{M_V^2}{F^2} \, .
\end{eqnarray}
In order to write the left-hand side of these equations we have employed the relations~(\ref{eq.dictionary}).
The relations involving the
$V\varphi\varphi\varphi$ couplings
(one vector field and three Goldstone fields)
are omitted since they are irrelevant for our discussion~\cite{Dumm:2009kj}.

\vspace*{0.25cm}
\subsection{$\tau\to \varphi^- \gamma\nu_\tau$
and $\pi^0\to \gamma^*\gamma^*$ form-factors}

The $\tau\to \varphi^- \gamma\nu_\tau$ decay
($\varphi^-=\pi^-,\,  K^-$) is described by a vector and an axial-vector
form-factors $F_V(t)$ and $F_A(t)$, respectively,
which were computed in the R$\chi$T framework in Ref.~\cite{Guo:2010dv}.
The requirement that $F_V(t)$ vanishes at high momentum transfer
($t\to \infty$)~\cite{Brodsky:1973kr}
produces the constraints
\bear
M_V(2 \kappa_{12}^V +\kappa_{16}^V - 2\kappa_{17}^V) \quad =&
c_1 \, - \, c_2 \, + \, c_5 \, & = \quad  0 \, ,
\label{eq.TFF-constraint1}
\\
-\, M_V\kappa_{17}^V \quad =& c_5-c_6 &=\quad
\Frac{N_C}{32\pi^2 } \Frac{M_V}{\sqrt{2} F_V}
\, +\, \Frac{F_V}{\sqrt{2} M_V} \, d_3\, ,
\label{eq.TFF-constraint2}
\eear
with $d_3=\kappa_3^{VV}$ in $\tilde{\mL}_{R\chi T}^{\rm odd}$ notation.

The $\pi^0(r)\to \gamma^*(p)\,\gamma^*(q)$ {TFF},
$\mF_{\pi\gamma^*\gamma^*}(p^2,q^2)$,
was studied in Ref.~\cite{Kampf:2011ty}
by means of the $\tilde{\mL}^{\rm odd}_{R\chi T}$ Lagrangian.
Requiring that $\mF_{\pi\gamma^*\gamma^*}(0,q^2)$,
with one on-shell photon, vanishes at high momentum transfer~\cite{Brodsky:1973kr},
yields precisely the two previous constraints in
eqs.~(\ref{eq.TFF-constraint1}) and (\ref{eq.TFF-constraint2}).
One reaches this result if no further short-distance constraints are applied
(like, for instance, those from the $\bra VVP\ket$).
Moreover, the requirement that $\mF_{\pi\gamma^*\gamma^*}(q^2,q^2)$, with both
photons off-shell, vanishes when $q^2\to \infty$~\cite{off-shell-TFF} yields the
additional relation
\bear
-\, M_V \kappa_{17}^V \quad = &  c_5 \, - \, c_6 \, & =
\quad  \frac{N_C \,M_V}{64 \,\sqrt{2}\, \pi^2\,F_V} \, .
\label{eq.TFF-constraint3}
\eear
Remarkably, although just the $\pi^0\to \gamma^*\gamma^*$ TFF
was constrained to achieve this equation, it reproduces exactly
the short-distance $\bra VVP \ket$ relation in eq.~(\ref{eq.constraint-Kampf-C}).

Ref.~\cite{Kampf:2011ty}, on the other hand, substituted the $\bra VVP\ket$
relations~(\ref{eq.constraint-Kampf-C})--(\ref{eq.constraint-Kampf-E}) and expressed
the $\mF_{\pi\gamma^*\gamma^*}(0,q^2)$ constraint in the form
\bear
\label{eq: relation pigammagamma Kampf}
 1  +  \Frac{32 \sqrt{2}  F_V d_m \kappa_3^{PV}}{F^2} &=& 0\,.
\eear

\vspace*{0.25cm}
\subsection{$\tau\to (\varphi V)^- \nu_\tau$ vector form-factor}

The transition $\tau\to (\varphi V)^- \nu_\tau$
(with $\varphi=\pi^{-,\, 0} ,\,   K^{-,\, 0}$
and $V=\rho^{0,\, -},\, \omega,\, \overline{K}^{*0},\, K^{*0,\, -}$)
is parametrized by one vector form-factor $V(t)$
and three axial-vector form-factors $A_{1,2,3}(t)$. It was computed
in Ref.~\cite{Guo:2008} by means
of R$\chi$T for the case with two vector resonance multiplets.
High-energy constraints where extracted after demanding that these form-factors
vanished for large momentum transfer~\cite{Brodsky:1973kr}.
Restricting ourselves to the scenario with only one vector resonance multiplet studied here
the vector form-factor relations turn into~\cite{Guo:2008}
\bear
M_V(2 \kappa_{12}^V +\kappa_{16}^V - 2\kappa_{17}^V) \quad =&
c_1 \, - \, c_2 \, + \, c_5 \, & = \quad  0 \, ,
\label{eq.VP-VFF1}
\\
-\, M_V\kappa_{17}^V \quad =& c_5-c_6 &=\quad
\, - \, \Frac{F_V}{\sqrt{2} M_V} \, d_3\, ,
\label{eq.VP-VFF2}
\eear
with $d_3=\kappa_3^{VV}$ in $\tilde{\mL}_{R\chi T}^{\rm odd}$ notation.

\vspace*{0.25cm}
\subsection{Compatibility between constraints}

In a first step, we find that
the three $\tau^-\to (KK\pi)^-\nu_\tau$ relations~(\ref{eq:1c})--(\ref{eq:3c})
are compatible with the $\bra VVP\ket $ relations in
eqs.~(\ref{eq.constraint-Kampf-A})--(\ref{eq.constraint-Kampf-E}) provided
\bear
 1  +  \Frac{32 \sqrt{2}  F_V d_m \kappa_3^{PV}}{F^2} &=& 0\,,
\label{eq.compatibility1}
\\
F_V^2 &=& 3 F^2\, .
\label{eq.compatibility2}
\eear
The first relation, eq.~(\ref{eq.compatibility1}),
was previously obtained in Ref.~\cite{Kampf:2011ty}
(eq.~(\ref{eq: relation pigammagamma Kampf})) after requiring the right
short-distance behaviour for both the $\bra VVP\ket$ Green function and
the $\pi^0\to\gamma\gamma^*$ {TFF}.
This condition obviously requires $\kappa_3^{PV}\neq 0$,
{\it i.e.}, the presence of a pseudoscalar resonance contribution.

The second relation, eq.~(\ref{eq.compatibility2}), was also found
in the study of the radiative $\tau \to \varphi^-\gamma \nu_\tau$
processes in Ref.~\cite{Guo:2010dv}.
Ref.~\cite{Dumm:2009kj} pointed out that,
while the $\bra VVP\ket$ constraints
(\ref{eq: constraints PedroA})--(\ref{eq: constraints PedroE})
without pseudoscalar resonances yielded a wrong high-energy
structure for $\pi^0\to\gamma\gamma^*$,
the $\tau\to (KK\pi)^-\nu_\tau$ conditions~(\ref{eq:1c})-(\ref{eq:3c})
ensured the proper Brodsky-Lepage
asymptotic behaviour for the $\pi^0\to\gamma\gamma^*$ TFF,
provided that
the constraint~(\ref{eq.compatibility2}) is fulfilled.

It is noteworthy that the conditions~(\ref{eq.compatibility1})
and~(\ref{eq.compatibility2})
exactly agree with the high-energy
constraints for the $\tau\to \varphi^-\gamma \nu_\tau$ vector form-factor
(eqs.~(\ref{eq.TFF-constraint1}) and~(\ref{eq.TFF-constraint2})),
$\pi^0\to\gamma^*\gamma^*$ {TFF}
(eqs.~(\ref{eq.TFF-constraint1})--(\ref{eq.TFF-constraint3}))
and the $\tau\to (\varphi\, V)^-\nu_\tau$ vector form-factor
(eqs.~(\ref{eq.VP-VFF1}) and~(\ref{eq.VP-VFF2})).

\vspace*{0.25cm}
\section{Discussion and comparison with the phenomenology}

In summary, we provide in the present letter
the unique set of consistent high-energy constraints
in the odd intrinsic-parity sector
\begin{eqnarray}
M_V( 2 \kappa_{12}^V + 4 \kappa_{14}^V +\kappa_{16}^V -\kappa_{17}^V) \quad =&
4 \,c_3 \,+\, c_1 & = \quad  0\,,
\nonumber
\\[3mm]
M_V (2\kappa_{12}^V + \kappa_{16}^V -2\kappa_{17}^V) \quad =&
c_1 \, - \, c_2 \, + \, c_5 \, & = \quad  0 \, ,
\nonumber
\\[3mm]
-\, M_V \kappa_{17}^V \quad = &  c_5 \, - \, c_6 \, & =
\quad  \frac{N_C \,M_V}{64 \,\sqrt{2}\, \pi^2\,F_V} \, ,
\nonumber
\\[3mm]
8\kappa_2^{VV} \quad =&  d_1 \, + \, 8 \,d_2 & = \quad
\frac{F^2}{8\,F_V^2} - \frac{N_C \,M_V^2}{64\, \pi^2\, F_V^2}\,,
\nonumber
\\[3mm]
\kappa_3^{VV}\quad = & d_3& =\quad  -\, \Frac{N_C}{64\pi^2} \Frac{M_V^2}{F_V^2} \,,
\nonumber
\\[3mm]
& 1 \, + \, \Frac{32 \,\sqrt{2} \, F_V \,d_m\, \kappa_3^{PV}}{F^2} & = \quad  0\,,
\nonumber
\\[3mm]
& F_V^2 & = \quad  3\, F^2\,,
\label{eq: Consistent set of relations}
\end{eqnarray}
compatible for the
$\bra VVP\ket$ Green function~\cite{RuizFemenia:2003hm, Kampf:2011ty}
and a series of related odd intrinsic-parity amplitudes: the $\tau\to X^- \nu_\tau$
vector form-factors
($X^-=(KK\pi)^-$~\cite{Dumm:2009kj}, $\varphi^-\gamma$~\cite{Guo:2010dv},
$(\varphi V)^-$~\cite{Guo:2008})
and the $\pi^0\to\gamma^*\gamma^*$ {TFF}~\cite{Kampf:2011ty}.
The consistent set of high-energy relations~(\ref{eq: Consistent set of relations})
for these anomalous QCD amplitudes
constitutes the central outcome of this letter.
Notice however that only the pseudo-Goldstones
and the lightest vector and pseudoscalar resonance multiplets have been
considered here, so these relations would change if we varied the resonance content
of the theory.

The relations~(\ref{eq: Consistent set of relations}) also
agree with the short-distance constraints obtained in the analysis of
other anomalous processes in the resonance region:
the $\tau^-\to\eta\pi^-\pi^0\nu_\tau$ decay~\cite{Dumm:2012vb};
$V\to \varphi\gamma^{(\star)}$ and
$\varphi\to\gamma\gamma^{(\star)}$~decays~\cite{Chen:2012vw};
holographic studies of three-point Green functions and associated
form-factors \cite{Colangelo:2012ipa}; $e^+e^-\to \varphi \pi^+\pi^-$
($\varphi=\pi^0,\, \eta$)~\cite{Dai:2013joa}; and
the $\tau^- \to\pi^-\nu_\tau \ell^+ \ell^-$~decay~\cite{Guevara:2013wwa}.

Interestingly, an analogous set of consistent relations
was extracted in the even intrinsic-parity sector in~Ref.~\cite{Cirigliano:2006hb}
(eqs. (4.1), (4.2), (5.7) and (5.12)) for the $\bra VAP\ket$ Green
function~\cite{Cirigliano:2004ue} and related form-factors~\cite{GomezDumm:2003ku,Dumm:2009va}, in combination
with the $\pi\pi$ vector form-factor constraint and the two $VV-AA$
WSRs~\cite{Ecker:1989yg}.

The last constraint in~(\ref{eq: Consistent set of relations}), $F_V^2 = 3 F^2$,
is particularly interesting, as it was also previously derived
in the even intrinsic-parity sector in an independent way.
It was found in the
high-energy analysis of the $\pi\pi$ vector form-factor
at NLO in $1/N_C$~\cite{L9-Rosell}  if the scalar resonance effects are disregarded. Likewise,
the combination of the large--$N_C$ constraints for
$\pi\pi$ vector form-factor ($F_V G_V=F^2$)~\cite{Ecker:1989yg}
and $\pi\pi$--scattering ($3 G_V^2=F^2$ if the scalar resonance contributions
are neglected)~\cite{Guo:2007} also reproduces the condition $F_V^2=3 F^2$.
These two relations from $\pi\pi$ VFF and scattering also show up
in the context of holographic models of QCD~\cite{Colangelo:2012ipa,holo-SR1,holo-SR2}
when the derived sum-rules are restricted to the single vector resonance approximation.
Indeed, recent studies in that field~\cite{Colangelo:2012ipa}
are pointing out an interconnection between the even intrinsic-parity and
anomalous sectors of QCD~\cite{Colangelo:2012ipa,holo-SR2,Son-Yamamoto}.
It is also worth to note that this value $F_V=\sqrt{3} F = 3 G_V$
was found to be a low-energy fixed point of the renormalized couplings
$F_V(\mu)$ and $G_V(\mu)$ within
the single vector resonance approximation,
being NLO corrections in $1/N_C$ of the order of
$\Gamma_\rho/M_\rho\sim 20\%$~\cite{RChT-RGE}.

The combined study of the $\pi\pi$ VFF and the
$\pi\to\gamma\ell\nu$ axial-vector form-factor produces
$F_V\,=\,\sqrt{2}\,F=2\,G_V$
if operators with two or more resonance fields are disregarded~\cite{Ecker:1989yg},
although this is no longer so when they are taken
into full consideration~\cite{Cirigliano:2004ue}.
The large--$N_C$ study of the $\pi\pi$ partial-wave scattering
amplitudes $T^I_J(s)$   at high energies~\footnote{
A  generalization of the KSRF relation
was derived in Ref.~\cite{Guo:2007}
by demanding that the $T^I_J(s)$ obeyed once-subtracted dispersion relations
at large $N_C$. The contribution from the lightest vector and scalar resonance
exchanges in the $s$, $t$ and $u$ channels were computed therein, leading to
the large-$N_C$  constraint given
in eq.~(\ref{Generalized relation}).}
yields the generalized version of the KSRF relation~\cite{KSRF} including
the effect of scalar resonances~\cite{Guo:2007},
\begin{equation}
\label{Generalized relation}
  2\,c_d^2\,+\,3\,G_V^2\,=\,F^2 \, .
\end{equation}
 Nonetheless, Ref.~\cite{NPR} observed that  this relation
left a residual logarithmically divergent behaviour in the
$T^I_J(s)$
at high energies.
Moreover, eq.~(\ref{Generalized relation}) is incompatible
with the $\pi\pi$ VFF constraint $F_VG_V=F^2$~\cite{Ecker:1989yg}
and the odd-sector relation $F_V^2=3 F^2$
in eq.~(\ref{eq: Consistent set of relations})
if the scalar is taken into account and only the lightest multiplets are accounted for.
On the other hand, phenomenologically, it is found to be
reasonably well fulfilled and  violations are mild:
the resonance couplings extracted from pion and kaon
scattering phase-shift fits to data
were found to fulfill eq.~(\ref{Generalized relation})
within $15\leftrightarrow20 \%$ violations~\cite{GO},
regardless of taking or not the $c_d\to0$ limit.
This slight tension
can be relaxed if one discards the $\pi\pi$ partial-wave
constraint in eq.~(\ref{Generalized relation})
when only the lowest resonance multiplets 
are   
included. Alternatively, Ref.~\cite{NPR} advocated that the forward
$\pi\pi$ scattering amplitude could be well
described at large $N_C$ with just the lightest vector and scalar,
obtaining the relation
$2 c_d^2 + G_V^2=F^2$.  
This stems from assuming that
the expected power behaviour from Regge theory
at high energies (or a less divergent one).  However, the latter constraint
in combination with the $\pi\pi$ VFF ($F_VG_V=F^2$) and odd-sector one
($F_V^2=3 F^2$) leads to a scalar coupling $c_d= F/\sqrt{3}\approx  50$~MeV,
which is far too large in comparison with previous phenomenological
determinations where $c_d\lsim  30$~MeV (see Refs.~\cite{GO,eta-etap}
and references therein).

A good probe of the $F_V^2 = 3 F^2$ relation
are the $\tau^-\to(\pi\pi\pi)^-\nu_\tau$ decays,
even though its check is not free of ambiguities related
to the treatment of the $\rho(1450)$ resonance or the $\pi\pi$
rescattering effects producing the $\sigma$ resonance.
In Ref.~\cite{Dumm:2009va} only the first contribution was included
(phenomenologically) and fitting the differential
decay widths as a function of the three-pion invariant masses yielded
a deviation of $\sim 13\%$ with respect to this prediction.
The updated R$\chi$T TAUOLA currents \cite{TAUOLA-RChL}
added also a modelization of the $\sigma$ effect and benefited from the
two-pion invariant mass distributions measured
by BaBar \cite{Ian} to reach good agreement with data \cite{Nugent:2013hxa}.
The fitted value of $F_V$ is consistent with the previous prediction at
one sigma, which is remarkable since
the quoted error is slightly below 5$\%$.

The consistent set of relations (\ref{eq: Consistent set of relations})
can also be tested through the lightest meson ($\pi^0/\eta/\eta^\prime$) TFF,
which are an essential ingredient for predicting the related pseudoscalar-exchange hadronic
light-by-light contribution to the muon anomalous magnetic moment.
Refs.~\cite{Kampf:2011ty,Roig:2014uja} obtain their best
agreement with the $\pi$ TFF data violating
only the prediction for $\kappa_3^{PV}$
in eqs.~(\ref{eq: Consistent set of relations}) by $4\leftrightarrow 6\%$.
The current understanding of
the $\eta-\eta^\prime$ mixing in the double-angle mixing scheme
allows the prediction of the related $\eta-\eta^\prime$ TFF \cite{Roig:2014uja},
which are found in good agreement with data as well.

We find, therefore, that the minimal hadronical ansatz \cite{MHA}
--consisting on including as many resonance multiplets
as needed to achieve a consistent set of
short-distance constraints--
reduces to the single (vector and pseudoscalar) resonance approximation for
the $\left\langle VVP \right\rangle$ Green function
and related anomalous form-factors.
One must be aware that the OPE constraints from the $\bra VVP\ket$ Green function
in Sec.~\ref{sec.VVP}, where the three four-momenta $p,q,r$
are taken to infinity  at the same rate,
depend  crucially on the inclusion of both the
lightest pseudoscalar and vector resonances.
For instance, the total absence of $P$~resonances leads to
roughly a factor 2 difference
with respect to alternative  determinations of $d_3$~\cite{4p-GF},
far more important than the impact of considering $F_V=\sqrt{2} F$
or $F_V=\sqrt{3} F$.
Obviously, the unique consistent set of short-distance
constraints (\ref{eq: Consistent set of relations}) would be modified
if the spectrum of the theory is enlarged.

Despite we deal with the odd-intrinsic parity sector,
the constraint $F_V=\sqrt{3} F$ belongs also to the normal parity sector,
where one can test the impact of heavier states. Therein,
the study of two-meson vector form factors
in hadronic tau decays has been sensitive to the effect of excited
resonances thanks to the very good quality data taken at B-factories.
In the discussion at the beginning of
Sec. 4 in Ref.~\cite{Guevara:2013wwa} (see also references therein)
it is seen that the modifications induced by the excited resonances
in the short-distance relations obtained
within the single resonance approximation are -at most- of $5\%$.

This observation cannot be \textit{a priori} generalized
to the asymptotic relations
involving couplings which describe interactions between more
than one resonance field,
where current data are not
precise enough to settle this issue, although large
deviations are not hinted by the time being.

In the anomaly sector, it is more difficult to quantify possible deviations
and the impact from heavier states, due to both the higher complexity of the asymptotic constraints
and the larger uncertainty of the measurements.
The addition of the first excited vector meson multiplet
does not change the analysis of asymptotic constraints derived from the
$V\varphi$ form factors \cite{Roig:2014uja}.
However, it does modify the high-energy restrictions from the radiative pion form factor
(see eqs. (A.9) and (A.10) in Ref.~\cite{Radpidec}).
According to~\cite{Radpidec},
the first two relations in our eq.~(\ref{eq: Consistent set of relations})
do not get modified by the inclusion of the first excited multiplet.

Nonetheless, the next three relations in~(\ref{eq: Consistent set of relations})
do indeed change.
For instance, the $(c_5-c_6)$ constraint becomes~\cite{Radpidec}
\begin{equation}\label{modified}
 (c_5 \, - \, c_6) \, +\, \frac{F_V' M_V}{F_V M_V'}(c_5'-c_6')\,
 = \,  \frac{N_C \,M_V}{64 \,\sqrt{2}\, \pi^2\,F_V}\,,
\end{equation}
where the primed parameters refer to the excited vector $V'$ and are defined in analogy to the respective $V$ couplings.

The second term on the left-hand side of
eq.~(\ref{modified}) modifies our original equation.
Apart from dynamical reasons which might suppress $c_5'-c_6'$
with respect to $c_5-c_6$, the factor
$\frac{F_V' M_V}{F_V M_V'}$
reduces the effect of the former coupling combination by $\sim 0.3$~\cite{Guo:2008}.
More complicated equations are found in~\cite{Radpidec}
involving $d_3$, $d_1+8d_2$ and analogous
and new excited resonances couplings.
Likewise, there is no significant tension between
the phenomenological analysis of
the $\tau \to \eta^{(')} \pi^-\pi^0\nu_\tau$  data~\cite{Dumm:2012vb}
and  the  short-distance constraints~(\ref{eq: Consistent set of relations}),
which allows us to extract a conservative estimate of the
impact of excited multiplets as less than $20\%$
(see Ref.~\cite{Dumm:2012vb} for details).

New, more precise measurements of hadronic (radiative) tau decays, $e^+e^-$ hadroproduction, vector meson decays, meson TFF and meson-meson scattering phaseshifts would be
extremely helpful tools in increasing our knowledge of the hadronization of QCD currents in its non-perturbative regime and, in particular, in ascertaining the role of
excited resonance multiplets in the corresponding dynamics and its effect in the short-distance relations obtained within the single resonance approximation. These are
expected at Belle-II and forthcoming facilities.

Finally, we want to call the attention of the reader to the relations in
eq.~(\ref{eq.dictionary}), which provide a dictionary between the two R$\chi$T
bases $\mL^{\rm odd}_{R\chi T}$~\cite{RuizFemenia:2003hm}
and $\tilde{\mL}^{\rm odd}_{R\chi T}$~\cite{Kampf:2011ty} that can be useful in future comparisons.

\vspace*{0.25cm}
\section*{Acknowledgements}
We wish to thank the hospitality of IFAE (Barcelona), where part of this work was done. This research has been financed by the Mexican and Spanish Governments and ERDF funds
from the European Union: Conacyt and DGAPA and grants PAPIIT IN106913, FPA2010-17747, FPA2011-25948, AIC-D-2011-0818, SEV-2012-0249 (Severo Ochoa Program),
CSD2007-00042 (Consolider Project CPAN)] and the Comunidad de Madrid [HEPHACOS S2009/ESP-1473].


\begin{thebibliography}{100}
\bibitem{Weinberg:1978kz}
  S.~Weinberg,
  Physica A {\bf 96} (1979) 327.

\bibitem{Gasser:1983yg}
  J.~Gasser and H.~Leutwyler,
  Annals Phys.\  {\bf 158} (1984) 142.

\bibitem{Gasser:1984gg}
  J.~Gasser and H.~Leutwyler,
  Nucl.\ Phys.\ B {\bf 250} (1985) 465.

\bibitem{Bijnens:1999sh}
  J.~Bijnens, G.~Colangelo and G.~Ecker,
  JHEP {\bf 9902} (1999) 020
  [hep-ph/9902437].

\bibitem{Bijnens:2001bb}
  J.~Bijnens, L.~Girlanda and P.~Talavera,
  Eur.\ Phys.\ J.\ C {\bf 23} (2002) 539
  [arXiv:hep-ph/0110400].

\bibitem{Wess:1971yu}
  J.~Wess and B.~Zumino,
  Phys.\ Lett.\ B {\bf 37} (1971) 95.

\bibitem{W}
  E.~Witten,
  Nucl.\ Phys.\ B {\bf 223} (1983) 422.

\bibitem{Ecker:1988te}
  G.~Ecker, J.~Gasser, A.~Pich and E.~de Rafael,
  Nucl.\ Phys.\ B {\bf 321} (1989) 311.

\bibitem{Ecker:1989yg}
  G.~Ecker, J.~Gasser, H.~Leutwyler, A.~Pich and E.~de Rafael,
  Phys.\ Lett.\ B {\bf 223} (1989) 425.

\bibitem{'tHooft:1973jz}
  G.~'t Hooft,
  Nucl.\ Phys.\ B {\bf 72} (1974) 461.

\bibitem{'tHooft:1974hx}
  G.~'t Hooft,
  Nucl.\ Phys.\ B {\bf 75} (1974) 461.

\bibitem{Witten:1979kh}
  E.~Witten,
  Nucl.\ Phys.\ B {\bf 160} (1979) 57.

\bibitem{truncation-Prades}
    J. Bijnens, E. Gamiz, E. Lipartia and J. Prades,
    JHEP {\bf 0304} (2003) 055
    [arXiv:hep-ph/0304222].

\bibitem{truncation}
    M. Golterman and S. Peris,
    Phys. Rev. D {\bf 74} (2006) 096002
    [arXiv:hep-ph/0607152];
%
    P. Masjuan and S. Peris,
    JHEP {\bf 0705} (2007) 040
    [arXiv:0704.1247 [hep-ph]].

\bibitem{Cirigliano:2006hb}
  V.~Cirigliano, G.~Ecker, M.~Eidem\"uller, R.~Kaiser, A.~Pich and J.~Portol\'es,
  Nucl.\ Phys.\ B {\bf 753} (2006) 139
  [arXiv:hep-ph/0603205].

 \bibitem{RuizFemenia:2003hm}
  P.~D.~Ruiz-Femen\'{\i}a, A.~Pich and J.~Portol\'es,
  JHEP {\bf 0307} (2003) 003
  [arXiv:hep-ph/0306157].

\bibitem{Kampf:2011ty}
  K.~Kampf and J.~Novotny,
  Phys.\ Rev.\ D {\bf 84} (2011) 014036
  [arXiv:1104.3137 [hep-ph]].

\bibitem{SVZ}
    M. A. Shifman, A. I. Vainshtein, and V. I. Zakharov,
    Nucl. Phys. B{\bf 147} (1979) 385;       
%
    B {\bf 147} (1979) 448.       

\bibitem{WeinbergSRS}
S.~Weinberg,
  Phys.\ Rev.\ Lett.\  {\bf 18} (1967) 507.

\bibitem{Knecht:2001xc}
  M.~Knecht and A.~Nyffeler,
  Eur.\ Phys.\ J.\ C {\bf 21} (2001) 659
  [arXiv:hep-ph/0106034].

\bibitem{EarlierWorks}
  B.~Moussallam,
  Phys.\ Rev.\ D {\bf 51} (1995) 4939
  [arXiv:hep-ph/9407402];
  Nucl.\ Phys.\ B {\bf 504} (1997) 381
  [arXiv:hep-ph/9701400].


\bibitem{Cirigliano:2004ue}
  V.~Cirigliano, G.~Ecker, M.~Eidem\"uller, A.~Pich and J.~Portol\'es,
  Phys.\ Lett.\ B {\bf 596} (2004) 96
  [arXiv:hep-ph/0404004].

\bibitem{Cirigliano:2005xn}
  V.~Cirigliano, G.~Ecker, M.~Eidem\"uller, R.~Kaiser, A.~Pich and J.~Portol\'es,
  JHEP {\bf 0504} (2005) 006
  [arXiv:hep-ph/0503108].

\bibitem{Mateu:2008}
    Matthias Jamin and  Vicent Mateu,
    JHEP 0804 (2008) 040
    [arXiv:0802.2669 [hep-ph]].

\bibitem{Brodsky:1973kr}
  S.~J.~Brodsky and G.~R.~Farrar,
  Phys.\ Rev.\ Lett.\  {\bf 31} (1973) 1153;
%
  G.~P.~Lepage and S.~J.~Brodsky,
  Phys.\ Rev.\ D {\bf 22} (1980) 2157.

\bibitem{Dumm:2009kj}
  D.~G\'omez Dumm, P.~Roig, A.~Pich and J.~Portol\'es,
  Phys.\ Rev.\ D {\bf 81} (2010) 034031
  [arXiv:0911.2640 [hep-ph]].

\bibitem{Floratos:1978jb}
  E.~G.~Floratos, S.~Narison and E.~de Rafael,
  Nucl.\ Phys.\ B {\bf 155} (1979) 115.

\bibitem{Guo:2010dv}
  Z.~-H.~Guo and P.~Roig,
  Phys.\ Rev.\ D {\bf 82} (2010) 113016
  [arXiv:1009.2542 [hep-ph]].

\bibitem{off-shell-TFF}
    Fred Jegerlehner and  Andreas Nyffeler,
    Phys.Rept. {\bf 477} (2009) 1   
    [arXiv:0902.3360 [hep-ph]];
%
    Edward V. Shuryak and A.I. Vainshtein,
    Nucl.Phys. B {\bf 199} (1982) 451.

\bibitem{Guo:2008} 	
    Zhi-Hui Guo,
    Phys.Rev. D {\bf 78} (2008) 033004
    [arXiv:0806.4322 [hep-ph]].

\bibitem{Dumm:2012vb}
  D.~G\'omez Dumm and P.~Roig,
  Phys.\ Rev.\ D {\bf 86} (2012) 076009
  [arXiv:1208.1212 [hep-ph]].

\bibitem{Chen:2012vw}
  Y.~-H.~Chen, Z.~-H.~Guo and H.~-Q.~Zheng,
  Phys.\ Rev.\ D {\bf 85} (2012) 054018
  [arXiv:1201.2135 [hep-ph]].

\bibitem{Colangelo:2012ipa}
  P.~Colangelo, J.~J.~Sanz-Cillero and F.~Zuo,
  JHEP {\bf 1211} (2012) 012
  [arXiv:1207.5744 [hep-ph]];
%
    JHEP  {\bf 1306} (2013) 020
    [arXiv:1304.3618 [hep-ph]].

\bibitem{Dai:2013joa}
  L.~Y.~Dai, J.~Portol\'es and O.~Shekhovtsova,
  Phys.\ Rev.\ D {\bf 88} (2013) 056001
  [arXiv:1305.5751 [hep-ph]].

\bibitem{Guevara:2013wwa}
  P.~Roig, A.~Guevara and G.~L\'opez Castro,
  Phys.\ Rev.\ D {\bf 88} (2013) 033007
  [arXiv:1306.1732 [hep-ph]].

\bibitem{GomezDumm:2003ku}
 D.~G\'omez Dumm, A.~Pich and J.~Portol\'es,
 Phys.\ Rev.\ D {\bf 69} (2004) 073002
 [arXiv:hep-ph/0312183].

\bibitem{Dumm:2009va}
  D.~G\'omez Dumm, P.~Roig, A.~Pich and J.~Portol\'es,
  Phys.\ Lett.\ B {\bf 685} (2010) 158
  [arXiv:0911.4436 [hep-ph]].

\bibitem{MHA}
M.~Knecht, S.~Peris and E.~de Rafael,
Phys.\ Lett.\ B {\bf 443} (1998) 255
[arXiv:hep-ph/9809594];
 B {\bf 457} (1999) 227
 [arXiv:hep-ph/9812471];
 S.~Peris, M.~Perrottet and E.~de Rafael,
    JHEP {\bf 9805} (1998) 011
 [arXiv:hep-ph/9805442];
M.~F.~L.~Golterman and S.~Peris,
  Phys.\ Rev.\ D {\bf 61} (2000) 034018.
  M.~Knecht, S.~Peris, M.~Perrottet and E.~de Rafael,
  Phys.\ Rev.\ Lett.\  {\bf 83} (1999) 5230
  [arXiv:hep-ph/9908283].

\bibitem{L9-Rosell}
    A. Pich, I. Rosell and  J.J. Sanz-Cillero,
    JHEP 1102 (2011) 109
    [arXiv:1011.5771 [hep-ph]].

\bibitem{Guo:2007}
    Z.H. Guo, J.J. Sanz Cillero and  H.Q. Zheng,
    JHEP 0706 (2007) 030
    [arXiv:hep-ph/0701232].

\bibitem{holo-SR1}
    J. Hirn and V. Sanz,
    JHEP {\bf 12} (2005) 030
    [arXiv:hep-ph/0507049];
%
    R.S. Chivukula, M. Kurachi and M. Tanabashi,
    JHEP {\bf 06} (2004) 004
    [arXiv:hep-ph/0403112].

\bibitem{holo-SR2}
    T. Sakai and S. Sugimoto,
    Prog. Theor. Phys. {\bf 114} (2005) 1083
    [arXiv:hep-th/0507073].

\bibitem{Son-Yamamoto}
    Dam T. Son and Naoki Yamamoto,
    [arXiv:1010.0718 [hep-ph]].

\bibitem{KSRF}
    K. Kawarabayashi and M. Suzuki,
    Phys. Rev. Lett. {\bf 16}   (1966) 255;
%
    Riazuddin and Fayazuddin,
    Phys. Rev. {\bf 147} (1966) 1071.

\bibitem{NPR}
  J.~Nieves, A.~Pich and E.~Ruiz Arriola,
  Phys.\ Rev.\ D {\bf 84} (2011) 096002
  [arXiv:1107.3247 [hep-ph]].

\bibitem{GO}
Z.~-H.~Guo and J.~A.~Oller,
  Phys.\ Rev.\ D {\bf 84} (2011) 034005
  [arXiv:1104.2849 [hep-ph]];
  \\
%
 Z.~-H.~Guo, J.~A.~Oller and J.~Ruiz de Elvira,
  Phys.\ Rev.\ D {\bf 86} (2012) 054006
  [arXiv:1206.4163 [hep-ph]].

\bibitem{eta-etap}
    R. Escribano,   P. Masjuan and  J.J. Sanz-Cillero,
    JHEP {\bf 1105} (2011) 094
    [arXiv:1011.5884 [hep-ph]].

\bibitem{TAUOLA-RChL}
  O.~Shekhovtsova, T.~Przedzinski, P.~Roig and Z.~Was,
  Phys.\ Rev.\ D {\bf 86} (2012) 113008
  [arXiv:1203.3955 [hep-ph]].

\bibitem{Ian}
  I.~M.~Nugent,
  arXiv:1301.7105 [hep-ex].
To be published in the Proceedings of the 12th International Workshop on Tau
Lepton Physics (TAU 2012).

\bibitem{4p-GF}
    B. Ananthanarayan and  B. Moussallam,
    JHEP {\bf 0406} (2004) 047
    [arXiv:hep-ph/0405206].

\bibitem{Nugent:2013hxa}
  I.~M.~Nugent, T.~Przedzinski, P.~Roig, O.~Shekhovtsova and Z.~Was,
  Phys.\ Rev.\ D {\bf 88} (2013) 093012
  [arXiv:1310.1053 [hep-ph]].

\bibitem{Roig:2014uja}
  P.~Roig, A.~Guevara and G.~L\'opez Castro,
  arXiv:1401.4099 [hep-ph], to appear in Phys.\ Rev.\ D.

\bibitem{Radpidec}
 V.~Mateu and J.~Portol\'es,
 Eur.\ Phys.\ J.\ C {\bf 52} (2007) 325
 [arXiv:0706.1039 [hep-ph]].

\bibitem{RChT-RGE}
    J.J. Sanz-Cillero,
    Phys.Lett. B {\bf 681} (2009) 100   
    [arXiv:0905.3676 [hep-ph]].

\end{thebibliography}
\end{document}